\documentclass[aps,prl,showpacs,twocolumn]{revtex4}
\usepackage{graphicx}
\usepackage{amsmath}
\usepackage{amsfonts}

\begin{document}

\title{Non-Abelian Quantum Hall Effect in Topological Flat Bands}
\author{Yi-Fei Wang$^{1,2}$, Hong Yao$^{3}$, Zheng-Cheng Gu$^{4}$,
Chang-De Gong$^{1,5}$, and D. N. Sheng$^{2}$}
\affiliation{$^1$Center for Statistical and Theoretical Condensed
Matter Physics, and Department of Physics, Zhejiang Normal
University, Jinhua 321004, China \\$^2$Department of Physics and
Astronomy, California State University, Northridge, California
91330, USA \\$^3$Department of Physics, Stanford University,
Stanford, California 94305, USA
\\$^4$Kavli Institute for Theoretical Physics, University
of California, Santa Barbara, California 93106, USA
\\$^5$National Laboratory of Solid State Microstructures
and Department of Physics, Nanjing University, Nanjing 210093,
China}
\date{\today}

\begin{abstract}
Inspired by recent theoretical discovery of robust fractional
topological phases without a magnetic field, we search for the
non-Abelian quantum Hall effect (NA-QHE) in lattice models with
topological flat bands (TFBs). Through extensive numerical studies
on the Haldane model with three-body hard-core bosons loaded into a
TFB, we find convincing numerical evidence of a stable $\nu=1$
bosonic NA-QHE, with the characteristic three-fold quasi-degeneracy
of ground states on a torus, a quantized Chern number, and a robust
spectrum gap. Moreover, the spectrum for two-quasihole states also
shows a finite energy gap, with the number of states in the lower
energy sector satisfying the same counting rule as the Moore-Read
Pfaffian state.

\end{abstract}

  \maketitle

{\it Introduction.---}The fractional quantum Hall effect (FQHE) is
one of the most fascinating phenomena in interacting quantum
many-particle systems. While many Abelian quantum Hall states have
been discovered in a two-dimensional electron gas, much effort has
been devoted to studies of the non-Abelian quantum Hall effect
(NA-QHE) since it was proposed two decades
ago~\cite{Moore,Greiter,NA_Wen,Read,Nayak,Morf,haldane52,geometry,DHLee,Levin}.
One promising experimental candidate for the NA-QHE is the $\nu=5/2$
state with electrons occupying the second Landau level
(LL)~\cite{xia2004}, while the nature of other candidates like the
$\nu=12/5$ state remains less settled. In addition, the NA-QHE is
believed to be possible in fast rotating Bose-Einstein condensates
(BEC)~\cite{Cooper2,Regnault} and optical lattices with an
artificial gauge field~\cite{Cirac}. The $\nu=1$ bosonic Pfaffian
state found in the lowest LL~\cite{Cooper2,Regnault} is favored by a
three-body repulsive interaction~\cite{Greiter}. In a lattice model,
such repulsive interaction can be realized by imposing the
three-body hard-core boson constraint~\cite{Cirac} that no more than
two bosons on any site are allowed. Interestingly, the three-body
hard-core bosons can be mapped to spin $S=1$ systems, and a NA
chiral spin liquid wave function has been proposed
recently~\cite{Greiter2}. The NA-QHE not only is an interesting
many-body phenomenon, but also provides a promising platform for
implementing topological quantum computation~\cite{Nayak2}.

Recently, systematic numerical works demonstrated convincing
evidence of the Abelian FQHE of interacting fermions and hard-core
bosons~\cite{Sheng1,YFWang,Regnault2} in topological flat band (TFB)
models~\cite{Wen}. Such TFB models belong to the topological class
of the well-known Haldane model~\cite{Haldane} with at least one
topologically nontrivial nearly-flat band carrying a nonzero Chern
number, which is also separated from the other bands by large
gaps~\cite{Wen,Fiete,Venderbos,Mueller}. This intriguing
fractionalization effect in TFBs without LLs has stimulated a lot of
recent research
activities\cite{XLQi,Murthy,parton,Neupert,Xiao,Sondhi,Goerbig}. The
$\nu=1/2$ bosonic FQHE found in TFB models for hard-core
bosons~\cite{YFWang} can also be considered as one example of the
long-sought chiral spin states for spin $1/2$
system~\cite{Laughlin}. It is now tempting to take the TFB as a
promising testbed to search for more exotic quantum Hall states,
possibly with non-Abelian nature.

\begin{figure}[tb]
  \vspace{-0.1in}
  \hspace{0.0in}
 \includegraphics[scale=0.72]{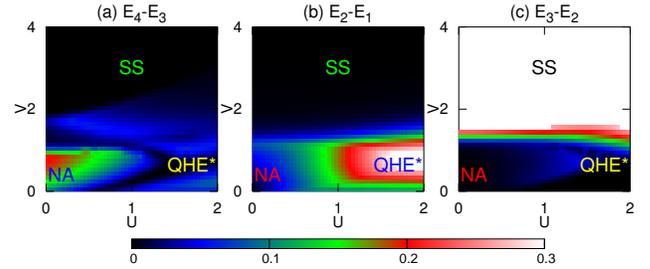}
  \vspace{-0.15in}
  \caption{(color online). Intensity plots of spectrum gaps in $U$-$V$ phase
  space for the $20$-site lattice at $\nu=1$.
   $E_1$, $E_2$, $E_3$ and $E_4$ denote the energies of the lowest four eigenstates.
   NA, QHE$^*$ and SS label the rough phase regions of NA-QHE, (non-degenerate)
   quantum Hall phase, and supersolid inferred from
   all three spectrum-gap plots and other information (see the text).} \label{f.1}
\end{figure}

In this letter, we search for the non-Abelian quantum phase for
bosons in TFBs without LLs. Through extensive exact diagonalization
(ED) study on the Haldane model with three-body hard-core bosons
loaded into a TFB, we find convincing numerical evidence of the
$\nu=1$ bosonic NA-QHE, with the characteristic three-fold
quasi-degeneracy of ground states (GSs) on a
torus~\cite{Greiter,Read,Cooper2}, an integer quantized Chern number
associated with GSs, and a robust spectrum gap in a finite region of
the parameter space. An energy gap is also found to separate the low
energy quasihole states from the higher energy ones, indicating the
existence of the ``zero-energy'' sector (for the interacting
Hamiltonian)~\cite{haldane52,haldane_donna}, as in the Moore-Read
Pfaffian state. The number of quasihole states in the lower energy
sector also satisfy the same counting rule as the Moore-Read
state~\cite{DHLee,haldane52,Regnault2,XGWen}. We further obtain the
quantum phase diagram based on our ED studies and illustrate quantum
phase transitions of the NA-QHE phase to other competing states.

{\it Formulation.---}We study the Haldane
model~\cite{Haldane} on the honeycomb lattice with interacting
bosons loaded into a TFB~\cite{YFWang}:
\begin{eqnarray}
H= &-&t^{\prime}\sum_{\langle\langle\mathbf{r}\mathbf{r}^{
\prime}\rangle\rangle}
\left[b^{\dagger}_{\mathbf{r}^{ \prime}}b_{\mathbf{r}}\exp\left(i\phi_{\mathbf{r}^{ \prime}\mathbf{r}}\right)+\textrm{H.c.}\right]\nonumber\\
&-&t\sum_{\langle\mathbf{r}\mathbf{r}^{ \prime}\rangle}
\left[b^{\dagger}_{\mathbf{r}^{\prime}}b_{\mathbf{r}}+\textrm{H.c.}\right]
-t^{\prime\prime}\sum_{\langle\langle\langle\mathbf{r}\mathbf{r}^{
\prime}\rangle\rangle\rangle}
\left[b^{\dagger}_{\mathbf{r}^{\prime}}b_{\mathbf{r}}+\textrm{H.c.}\right]\nonumber\\
&+&\frac{U}{2}\sum_{\mathbf{r}}
n_{\mathbf{r}}\left(n_{\mathbf{r}}-1\right)+V\sum_{\langle\mathbf{r}\mathbf{r}^{
\prime}\rangle} n_{\mathbf{r}}n_{\mathbf{r}^{\prime}} \label{e.1}
\end{eqnarray}
where $b^{\dagger}_{\mathbf{r}}$ creates a three-body hard-core
boson at site $\mathbf{r}$ satisfying
$\left(b^{\dagger}_{\mathbf{r}}\right)^3=0$ and
$\left(b_{\mathbf{r}}\right)^3=0$~\cite{Cirac}; $U$ and $V$ are the
two-body on-site and nearest-neighbor (NN) interactions. This model
can also be considered as a spin-1 model via the standard mapping
from the three-body hard-core bosons to the $S=1$ spins. It is clear
that $U/t\to\infty$ corresponds to the limit of the (two-body)
hard-core bosons.

The honeycomb lattice has a unit cell of two sites, and thus has two
single-particle bands. Here, we adopt the previous parameters $t=1$,
$t^{\prime}=0.60$, $t^{\prime\prime}=-0.58$ and $\phi/2\pi=0.2$,
such that a lower TFB is formed with a flatness ratio of about
$50$~\cite{YFWang}. For our ED study, we consider a finite system of
$N_1\times N_2$ unit cells (total number of sites $N_s=2N_1N_2$ and
total number of single-particle orbitals $N_{\rm orb}=N_1N_2$ in
each band) with basis vectors $\mathbf{a}_1$ and $\mathbf{a}_2$ and
periodic boundary conditions. We denote the boson numbers as $N_b$,
and the filling factor of the TFB is thus $\nu=N_b/N_{\rm orb}$. We
diagonalize the system Hamiltonian in each momentum
$\mathbf{q}=(2\pi k_1/N_1,2\pi k_2/N_2)$ sector, with $(k_1,k_2)$ as
integer quantum numbers.

{\it The phase diagram.---}We first look at the spectrum gaps for a
finite lattice with $N_s=20$ sites at filling $\nu=1$ as shown by
Fig.~\ref{f.1}, where $E_1$, $E_2$, $E_3$ and $E_4$ denote the
energies of the lowest four eigenstates, respectively. From the
three spectrum gaps $E_4-E_3$, $E_2-E_1$ and $E_3-E_2$, we can
obtain rather rich information of the possible phases and related
phase diagram. For the $\nu=1$ NA-QHE, two necessary conditions are:
a ground state manifold (GSM) with three quasi-degenerate lowest
eigenstates ($E_3-E_1\sim0$), and it is separated from the higher
eigenstates by a finite spectrum gap $E_4-E_3\gg E_3-E_1$. From the
Fig.~\ref{f.1}, it can be seen that both conditions are satisfied
simultaneously around the left bottom corner in the $U$-$V$ space.
The right bottom region is characterized by a finite $E_2-E_1$ gap
but a very small $E_3-E_2$, which is a possible QHE phase (labeled
as QHE$^*$ with more discussion to follow). While in the upper region
with larger $V$, the energy difference $E_2-E_1$ almost vanishes and a large
$E_3-E_2$ gap appears, indicating the two-fold quasi-degenerate
states. These are consistent with a sub-lattice solid order. Moreover, upon
changing boundary phases, the two lowest energy states evolve into
the higher energy spectrum, demonstrating its ``metallic'' nature in additional
to its ``solid'' feature. So we identify this phase as a supersolid (SS)
phase~\cite{MChan}. We have also obtained similar results from a
larger lattice with $N_s=24$ sites.

\begin{figure}[tb]
  \vspace{0.0in}
  \hspace{0.0in}
  \includegraphics[scale=0.58]{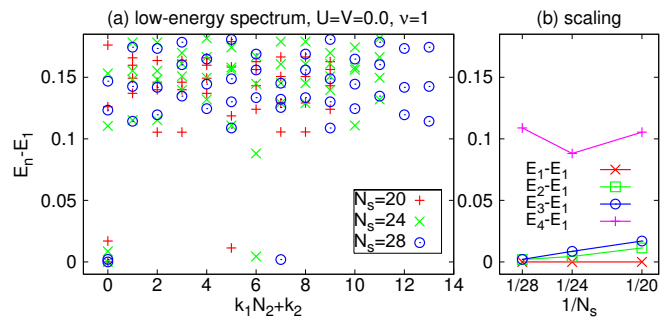}
  \vspace{-0.06in}
  \caption{(color online).
  (a) Low energy spectrum $E_n-E_1$ versus
  the momentum $k_1N_2+k_2$ of the NA-QHE phase for
  three lattice sizes $N_s=20$, $24$ and $28$.
  (b) Scaling plot of the spectrum gaps in (a).} \label{f.2}
\end{figure}

{\it Low energy spectrum.---}For the NA-QHE phase, we would like to
check whether the spectrum gap $E_4-E_3$ holds for other lattice
sizes. A few lowest states in each momentum sector of three system
sizes with $N_s=20$, $24$ and $28$ for the case of $U=V=0.0$ are
shown in Fig.~\ref{f.2}(a). The Hilbert subspaces of the $N_s=28$
lattice have the dimensions of about 700 million (which is about the
limit of the current ED method). We can see that, for each system
size, there is an obvious GSM with three-fold quasi-degenerate
states [two of them in the $(k_1,k_2)=(0,0)$ sector with very close
energies]. The GSM is well separated from the higher energy spectrum
by a large gap for all system sizes while the scaling plot of the
spectrum gap [Fig.~\ref{f.2}(b)]  suggests that the  gap $E_4-E_3$
and the three-fold quasi-degenerate GSM of the NA-QHE phase should
survive in the thermodynamic limit.

\begin{figure}[tb]
  \vspace{0.0in}
  \hspace{0.0in}
  \includegraphics[scale=0.58]{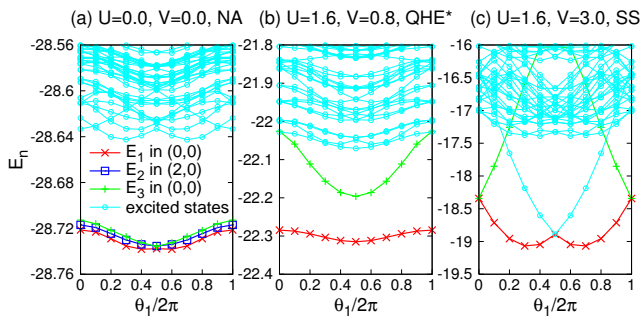}
  \vspace{-0.15in}
  \caption{(color online). Low energy spectra versus
  $\theta_1$ at a fixed $\theta_2=0$
  for three phases in the $N_s=24$ lattice at $\nu=1$:
  (a) the NA-QHE phase; (b) the QHE$^*$ phase; (c) the SS phase.} \label{f.3}
\end{figure}

\begin{figure}[tb]
  \vspace{-0.13in}
  \hspace{0.0in}
  \includegraphics[scale=0.62]{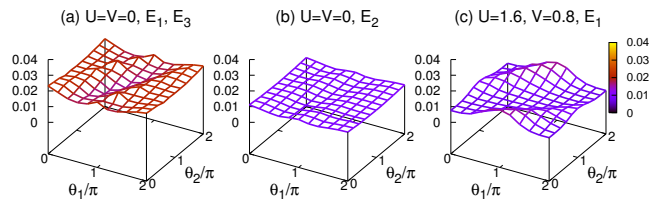}
  \vspace{-0.20in}
  \caption{(color online). Berry curvatures
  $F(\theta_1,\theta_2)\Delta\theta_1\Delta\theta_2/2\pi$
   at $10\times10$ mesh points for the $N_s=20$ cases:
  (a) the 1st and 3rd GSs [(0,0) sector] in the GSM of the NA-QHE phase;
  (b) the 2nd GS [(1,0) sector] in the GSM of the NA-QHE phase;
  (c) the single GS of the QHE$^*$ phase.} \label{f.4}
\end{figure}

{\it Berry curvature and Chern number.---}The Chern
number~\cite{Thouless} (i.e. the Berry phase in units of $2\pi$) of
a many-body state is an integral invariant in the boundary phase
space~\cite{Niu,Sheng2}: $C={{1}\over{2\pi}}\int d\theta_1 d\theta_2
F(\theta_1,\theta_2)$, where two boundary phases $\theta_1$ and
$\theta_2$ are introduced for the generalized boundary conditions in
$\mathbf{a}_1$ and $\mathbf{a}_2$ directions, and the Berry
curvature is given by $F(\theta_1,\theta_2)=\rm{Im}
\left(\left\langle {{\partial
\Psi}\over{\partial\theta_2}}\Big{|}{{\partial
\Psi}\over{\partial\theta_1}}\right\rangle -\left\langle {{\partial
\Psi}\over{\partial\theta_1}}\Big{|}{{\partial
\Psi}\over{\partial\theta_2}}\right\rangle\right)$. For the GSM of
NA-QHE phase with $N_s=24$, the three GSs maintain their
quasi-degeneracy and are well separated from the other low-energy
excitation spectrum when tuning the boundary phases, indicating the
robustness of the NA-QHE phase [Fig.~\ref{f.3}(a)]. Moreover, the
GSM in the NA-QHE phase shares a total Chern number $C=3$: e.g. for
the $N_s=20$ cases, we have two GSs of the GSM in the (0,0) sector
which contribute the integral Berry phase $4\pi$
[Fig.~\ref{f.4}(a)], the other GS of the GSM in the (1,0) sector
which contributes the integral Berry phase $2\pi$
[Fig.~\ref{f.4}(b)], and thus the total Chern number of the GSM is
$C=3$. In the possible QHE$^*$ phase, the single GS is well
separated from other low-energy excitation spectrum when tuning the
boundary phases [Fig.~\ref{f.3}(b)]. The Berry curvature of the
$N_s=20$ case is shown in Fig.~\ref{f.4}(c), which gives rise to  a
quantized Chern number $C=1$. On the other hand, for the SS phase,
the initial two-fold GS quasi-degeneracy is immediately destroyed
when tuning the boundary phases, the two GSs evolve into the higher
excitation spectrum. They do not have well-defined Chern numbers
since there is no well defined spectrum gap, which indicates a
``metallic'' feature of this SS phase [Fig.~\ref{f.3}(c)].

\begin{figure}[tb]
  \vspace{0.0in}
  \hspace{0.0in}
  \includegraphics[scale=0.58]{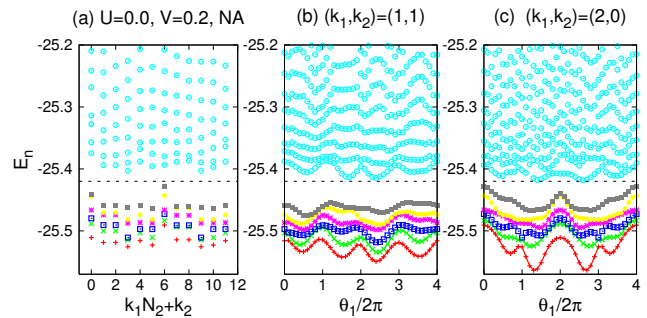}
  \vspace{-0.15in}
  \caption{(color online). (a) Quasihole excitation spectrum
  in the NA-QHE phase for $N_s=24$ and $N_b=11$. (b)-(c) Low energy spectra versus
  $\theta_1$ at a fixed $\theta_2=0$ in two momentum sectors of (a). } \label{f.5}
\end{figure}

{\it Quasihole excitation spectrum.---}In order to investigate the
possible fractional statistics of the NA-QHE state, we study the
quasihole excitation spectrum by removing one boson from the
$\nu=1$, and  expect two quasiholes of fractional bosonic charge
$1/2$~\cite{Moore,Greiter,Read}. As shown in Fig.~\ref{f.5}(a), for
a typical NA-QHE state on the $N_s=24$ lattice, the quasihole
spectrum exhibits a distinguishable gap which separates a few lowest
states in each momentum sector from the other higher-energy states.
For each momentum sector, we check the spectrum gap upon changing
boundary phases. For the sectors already with a large gap [e.g. the
$(1,1)$ sector], the spectrum gap is maintained well for all
boundary phases with 6 lowest states below the gap as shown in
Fig.~\ref{f.5}(b). For other sectors [e.g. the $(2,0)$ sector] where
the quasihole gap is less obvious,
 upon changing boundary phases the spectrum gap
becomes clearer and we also find 6 lowest states below the gap
as shown in Fig.~\ref{f.5}(c). By summing up all 12 sectors together, we have
72 low energy quasihole states in total. We also find similar
features for the $N_s=20$ (and $N_b=9$) case: there are 5 quasihole
states in each momentum sector, and all 10 sectors give 50 low
energy quasihole states in total.

\begin{figure}[tb]
  \vspace{0.0in}
  \hspace{0.0in}
  \includegraphics[scale=0.20]{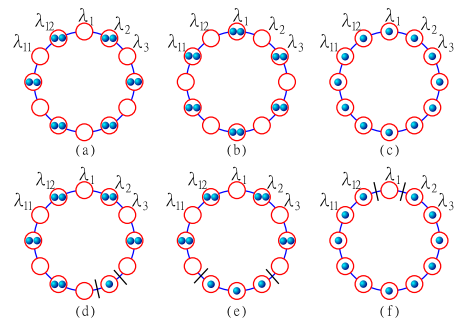}
  \vspace{-0.05in}
  \caption{(color online). Root configurations in
  the $N_{\rm orb}=12$ single-particle orbitals:
  (a)-(c) three GSs (02), (20) and (11);
  (d)-(f) two-quasihole states with two domain walls (represented by two vertical lines).} \label{f.6}
\end{figure}

The number of low energy two-quasihole states in the NA-QHE phase
described above can be heuristically understood from the counting
rule based on the generalized Pauli
principle~\cite{DHLee,Regnault2}. Using the Wannier representation
for a TFB~\cite{XLQi}, a set of $N_{\rm orb}=N_s/2$ periodic
single-particle orbitals are formed. Now, let us consider
a system with $N_{\rm orb}=12$  as an example. The generalized Pauli
principle that no more than two bosons occupying two consecutive
orbitals~\cite{DHLee,haldane52,Regnault2} results in only three GS root
configurations in the above orbitals
$|n_{\lambda_1},n_{\lambda_2},\dots, n_{\lambda_{N_{\rm
orb}}}\rangle$: $(02)\equiv|020202020202\rangle$,
$(20)\equiv|202020202020\rangle$, and
$(11)\equiv|111111111111\rangle$~[Figs.~\ref{f.6}(a)-(c)]. We now
count how many ways we can remove one boson from the above
three GS configurations $(02)$, $(20)$ and $(11)$. The boson
occupancy of two-quasihole states should be a mixture of two
segments of the three GS configurations, with two domain walls each
representing one quasihole with $1/2$ charge~\cite{DHLee}. A
simple analysis gives $6$ types of configurations with odd number of
$1's$: $\left|\dots 20|1|020\dots\right\rangle$, $|\dots
20|111|020\dots\rangle$, $|\dots 20|11111|020\dots\rangle$,
$\dots\dots$, and $|0|11111111111|\rangle$~[Figs.~\ref{f.6}(d)-(f)],
where two domain walls (quasiholes) are displayed by two vertical
lines $|$'s. Considering $12$ translations of the above $6$ states,
we finally get the total $72$ ($N^2_{\rm orb}/2$ in general)
two-quasihole states in exact accordance with our numerical
results.

\begin{figure}[tb]
  \vspace{0.0in}
  \hspace{0.0in}
  \includegraphics[scale=0.58]{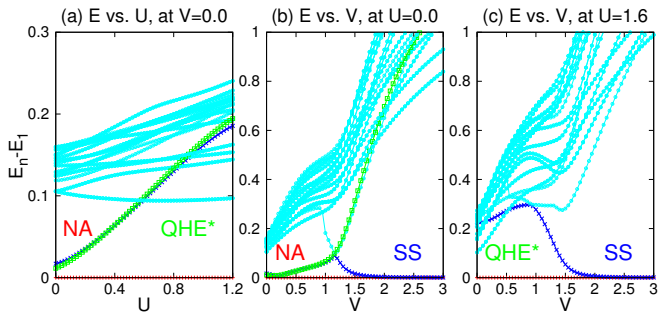}
  \vspace{-0.15in}
  \caption{(color online). Quantum phase transitions when tuning $U$ (or $V$) with
  $V$ (or $U$) fixed for $N_s=20$: (a) NA-QHE to QHE$^*$; (b) NA-QHE to SS; (c) QHE$^*$ to SS.} \label{f.7}
\end{figure}

We would like to emphasize that, for the GS obtained numerically,
the above mentioned root configurations are not the dominant
configurations due to quantum fluctuations just like the other FQHE
states in LLs on a torus. But the number of the low energy sector
quasihole states in QHE systems is a consequence of the Pauli
principle~\cite{DHLee,haldane52,Regnault2} resulting from strong
short-range pseudopotentials~\cite{haldane_donna,geometry}. Indeed,
based on the pattern of zeros classification  for FQHE states, the
Moore-Read state has a lower energy sector with the same number of
quasiholes as obtained in our numerical results~\cite{XGWen}.

{\it Quantum phase transitions.---}When tuning $U$ (or $V$) with
fixed $V$ (or $U$) away from the NA-QHE region, we observe quantum
phase transitions from the NA-QHE to other quantum phases including
a possible QHE$^*$ and the SS as shown by Fig.~\ref{f.7}.

Even though the NA-QHE phase is shown to be remarkably robust in our
ED study, we are less certain about the QHE$^*$ phase. As shown in
Fig.~\ref{f.7}(a), if we increase $U$, two higher energy states from
the GSM of the NA-QHE phase will emerge into the excited spectrum,
while the evolution of the wave function of the lowest energy state
is very smooth. Thus we suspect that the system may have a very long
correlation length near the possible QHE$^*$ phase, and we
conjecture that the spectrum gap above the GS may collapse when the
system size becomes very large. We leave this issue to be addressed
in future studies.

The SS phase has a two-fold GS quasi-degeneracy, strong
intra-sublattice density-density correlations and vanishing
inter-sublattice density-density correlations. These observations
indicate that the bosons prefer occupying one of the two sublattices.
By changing boundary phases, the two GSs of the SS state evolve into
the higher energy spectrum, indicating its ``metallic'' feature
besides its solid feature. However if we fix the $V$ as $3.0$ and go
to a larger $U>8.0$ (not shown in the phase diagram), the
``metallic'' character of the SS phase disappears  while the
solid feature remains indicating the system enters a pure solid
phase.

{\it Summary and discussion.---}The $\nu=1$ NA-QHE is characterized
by a three-fold quasi-degenerate GSM and a total Chern number $C=3$
carried by these states. The three-fold quasi-degenerate GSM is only
observed for system with even number of bosons ($N_b=10,12,14$) at
$\nu=1$, while it is absent for odd numbers of bosons (e.g. $N_b=9$
and $N_s=18$) consistent with the pairing nature for
Pfaffian-like states. Interestingly, the $\nu=1$ NA-QHE is already
quite stable with a clear GSM and a large spectrum gap for
three-body hard-core bosons without additional interactions
($U=V=0$). The spectrum gap can even be significantly enhanced with
the presence of a small $V$, which makes it possible for such a
state to be realized using optical lattices.

We thank Matthew Fisher, Hong-Chen Jiang, Steve Kivelson, Dung-Hai
Lee, and Michael Levin for helpful discussions. DNS thanks Duncan
Haldane for stimulating discussions. This work is supported by the
US DOE Office of Basic Energy Sciences under Grant No.
DE-FG02-06ER46305 (DNS), the NSFC of China Grant No. 10904130 (YFW),
the US DOE Grant No. DE-AC02-05CH11231 (HY), the US NSF Grant No.
NSFPHY05-51164 (ZCG), and the State Key Program for
Basic Researches of China Grants No. 2006CB921802
and No. 2009CB929504 (CDG).\\

{\it Note added.---}After the completion of this work, we became
aware of a related work by Bernevig and Regnault~\cite{Bernevig}
addressing the NA-QHE in fermionic TFB systems.


\begin{thebibliography}{99}

\bibitem{Moore} G. Moore and N. Read,
Nucl. Phys. B {\bf 360}, 362 (1991).
\bibitem{Greiter} M. Greiter, X. G. Wen, and F. Wilczek,
Phys. Rev. Lett. {\bf 66}, 3205 (1991); Nucl. Phys. B {\bf 374}, 567
(1992).

\bibitem{NA_Wen} X. G. Wen, Phys. Rev. Lett. {\bf 66}, 802 (1991);
Phys. Rev. Lett. {\bf 70}, 355 (1993).


\bibitem{Read} N. Read and E. Rezayi, Phys. Rev. B {\bf 54}, 16864 (1996);
N. Read and E. Rezayi, Phys. Rev. B {\bf 59}, 8084 (1999);
N. Read and D. Green, Phys. Rev. B {\bf 61}, 10267 (2000).

\bibitem{Nayak} C. Nayak and F. Wilczek, Nucl. Phys. B {\bf 479}, 529 (1996).

\bibitem{Morf} R. H. Morf, Phys. Rev. Lett. {\bf 80}, 1505 (1998);
E. H. Rezayi and F. D. M. Haldane, Phys. Rev. Lett. {\bf 84},
4685 (2000).


\bibitem{DHLee} A. Seidel and D. H. Lee,  Phys. Rev. Lett. {\bf 97}, 056804
(2006); E. J. Bergholtz, J. Kailasvuori, E. Wikberg, T. H. Hansson,
and A. Karlhede, Phys. Rev. B {\bf 74}, 081308(R) (2006).


\bibitem{haldane52} B. A. Bernevig and F. D. M. Haldane, Phys. Rev. Lett.
 {\bf 100}, 246802 (2008); H. Li and F. D. M. Haldane,
Phys. Rev. Lett. {\bf 101}, 010504 (2008);
E. Prodan and F. D. M. Haldane, Phys. Rev. B {\bf 80}, 115121 (2009).

\bibitem{geometry} F. D. M. Haldane, Phys. Rev. Lett. {\bf 107}, 116801 (2011).


\bibitem{Levin} M. Levin, B. I. Halperin, and B. Rosenow,
Phys. Rev. Lett. {\bf 99}, 236806 (2007); S.-S. Lee, S. Ryu, C.
Nayak, and M.P.A. Fisher, Phys. Rev. Lett. {\bf 99}, 236807 (2007);
M. R. Peterson, Th. Jolicoeur, and S. Das Sarma, Phys. Rev. Lett.
{\bf 101}, 016807 (2008); M. R. Peterson, K. Park, and S. Das Sarma,
{\bf 101}, 156803 (2008); H. Wang, D. N. Sheng, and F. D. M.
Haldane, Phys. Rev. B {\bf 80}, 241311 (2009).

\bibitem{xia2004} R. Willett et al., Phys. Rev. Lett. {\bf 59}, 1776 (1987);
J. P. Eisenstein et al., Phys. Rev. Lett. {\bf 88}, 076801 (2002);
J. Xia et al., Phys. Rev. Lett. {\bf 93}, 176809 (2004).

\bibitem{Cooper2}N. R. Cooper, N. K. Wilkin, and J. M. F. Gunn, Phys. Rev. Lett. {\bf
87}, 120405 (2001).
\bibitem{Regnault} N. Regnault and Th. Jolicoeur, Phys. Rev. Lett. {\bf 91}, 030402
(2003).

\bibitem{Cirac} B. Paredes, T. Keilmann, and J. I. Cirac, Phys. Rev. A {\bf 75}, 053611
(2007); L. Mazza, M. Rizzi, M. Lewenstein, and J. I. Cirac, Phys.
Rev. A {\bf 82}, 043629 (2010).


\bibitem{Greiter2} M. Greiter and R. Thomale,  Phys. Rev. Lett. {\bf 102},
207203 (2009); B. Scharfenberger, R. Thomale and M. Greiter, Phys.
Rev. B {\bf 84}, 140404(R) (2011). 


\bibitem{Nayak2} C. Nayak, S. H. Simon, A. Stern, M. Freedman,
and S. Das Sarma, Rev. Mod. Phys. {\bf 80}, 1083 (2008).


\bibitem{Sheng1} D. N. Sheng, Z. C. Gu, K. Sun, and L. Sheng,
Nature Commun. {\bf 2}, 389 (2011). 
\bibitem{YFWang} Y. F. Wang, Z. C. Gu, C. D. Gong, and D. N. Sheng,
Phys. Rev. Lett. {\bf 107}, 146803 (2011). 
\bibitem{Regnault2} N. Regnault and B. A. Bernevig,
Phys. Rev. X {\bf 1}, 021014 (2011). 


\bibitem{Wen} E. Tang, J. W. Mei, and X. G. Wen,
Phys. Rev. Lett. {\bf 106}, 236802 (2011); K. Sun, Z. C. Gu, H.
Katsura, and S. Das Sarma, Phys. Rev. Lett. {\bf 106}, 236803
(2011); T. Neupert, L. Santos, C. Chamon, and C. Mudry, Phys. Rev.
Lett. {\bf 106}, 236804 (2011).

\bibitem{Haldane} F. D. M. Haldane, Phys. Rev. Lett. {\bf 61}, 2015 (1988).

\bibitem{Fiete} X. Hu, M. Kargarian, and G. A. Fiete,
Phys. Rev. B {\bf 84}, 155116 (2011). 
\bibitem{Venderbos} J. W. F. Venderbos, M. Daghofer, and J. van den Brink,
Phys. Rev. Lett. {\bf 107}, 116401 (2011). 

\bibitem{Mueller} E. Kapit and E. Mueller,
Phys. Rev. Lett. {\bf 105}, 215303 (2010).


\bibitem{XLQi} X. L. Qi, Phys. Rev. Lett. {\bf 107}, 126803 (2011). 
\bibitem{Murthy} G. Murthy and R. Shankar, arXiv:1108.5501.
\bibitem{parton} Y. M. Lu and Y. Ran, arXiv:1109.0226;
J. McGreevy, B. Swingle, and K.-A. Tran, arXiv:1109.1569; A. Vaezi,
arXiv:1105.0406; F. Yang, X. L. Qi, and H. Yao, to be published.
\bibitem{Neupert} T. Neupert, L. Santos, S. Ryu, C. Chamon,
and C. Mudry, Phys. Rev. B {\bf 84}, 165107 (2011). 
\bibitem{Xiao} D. Xiao, W. Zhu, Y. Ran, N. Nagaosa, and S. Okamoto,
Nature Commun. {\bf 2}, 596 (2011). 
\bibitem{Sondhi} S. A. Parameswaran, R. Roy, and S. L. Sondhi,
arXiv:1106.4025.
\bibitem{Goerbig} M. O. Goerbig, Eur. Phys. J. B 85, 15 (2012). 


\bibitem{Laughlin} V. Kalmeyer and R. B. Laughlin,
Phys. Rev. Lett. {\bf59}, 2095 (1987); X. G. Wen, F. Wilczek, and A.
Zee, Phys. Rev. B {\bf 39}, 11413 (1989); H. Yao and S. A. Kivelson,
Phys. Rev. Lett. {\bf 99}, 247203 (2007); J. W. Mei, E. Tang, and X.
G. Wen, arXiv:1102.2406.



\bibitem{haldane_donna} F. D. M. Haldane and D. N. Sheng,  to be published.

\bibitem{XGWen} X.-G. Wen and Z. Wang, Phys. Rev. B {\bf 77}, 235108 (2008);
X.-G. Wen and Z. Wang, Phys. Rev. B {\bf 78}, 155109 (2008)


\bibitem{MChan}E. Kim and M. H. W. Chan, Nature (London) {\bf 427}, 225 (2004);
Science {\bf 305}, 1941 (2004).


\bibitem{Thouless} D. J. Thouless, M. Kohmoto, M. P. Nightingale, and M. den Nijs, Phys. Rev. Lett. {\bf 49}, 405
(1982).
\bibitem{Niu} Q. Niu, D. J. Thouless, and Y. S. Wu, Phys. Rev. B {\bf 31}, 3372 (1985).
\bibitem{Sheng2} D. N. Sheng, X. Wan, E. H. Rezayi, K. Yang, R. N. Bhatt, and F. D.
M. Haldane, Phys. Rev. Lett. {\bf 90}, 256802 (2003).


\bibitem{Bernevig} B. A. Bernevig and N. Regnault, arxiv:1110.4488.

\end{thebibliography}
\end{document}